\documentclass[12pt]{article}
\usepackage{psfig}
\usepackage{amssymb}
\begin{document}

\begin{center}
{\Large \bf Fluctuations near the deconfinement phase transition boundary}\\
\vspace{0.5cm}
{\large I.N. Mishustin \\
\vspace{0.5cm} 
{\it Frankfurt Institute for Advanced Studies, J.-W. Goethe University,
Max von Laue Str. 1, D-60438 Frankfurt am Main, Germany}\\
{\it The Kurchatov Institute, Russian Research Center, 123182 Moscow, Russia}}
\end{center}

\begin{center}
{\bf Abstract}
\end{center}
\medskip
In this talk I discuss how a first order phase transition may proceed in rapidly
expanding partonic matter produced in a relativistic heavy-ion collision.
The resulting picture is that a strong collective flow of matter will lead to
the fragmentation of a metastable phase into droplets. If the transition from
quark-gluon plasma to hadron gas is of the first order, it will manifest itself
by strong nonstatistical fluctuations in observable hadron distributions.
I discuss shortly existing experimental data on the multiplicity fluctuations.

\section{Introduction}

A general goal of present and future experiments with heavy-ion beams is to study
the properties of strongly interacting matter away from the nuclear ground
state. The main interest is focussed on searching for possible
phase transitions. Several phase transitions are predicted in different domains
of temperature $T$ and baryon density $\rho_B$. As well known,
strongly interacting matter has at least one multi-baryon bound state
at $\rho_B=\rho_0\approx0.16$ fm$^{-3}$ corresponding to normal nuclei.
It follows from the very existence of this bound state
that there should be a first order phase
transition of the liquid-gas type in normal nuclear matter at subsaturation
densities, $\rho_B<\rho_0$, and low temperatures, $T\leq 10$ MeV.
This phase transition manifests itself in a remarkable phenomenon
known as nuclear multifragmentation.

The situation at high $T$ and nonzero baryon chemical potential $\mu_B$
($\rho_B>0$) is not so clear, although everybody is
sure that the deconfinement and chiral transitions should occur somewhere.
Reliable lattice
calculations exist only for $\mu=0$ i.e. $\rho_B=0$ where they predict a
smooth deconfinement transition (crossover) at $T\approx 170$ MeV.
As model calculations show, the phase diagram in the
$(T,\mu_B)$ plane may contain a first order transition line (below called
the critical line) which ends at a (tri)critical point
\cite{Jac,Raj,Sca1}.  Possible signatures of this point in heavy-ion
collisions are discussed in ref. \cite{Ste}. However, it is unclear at present
whether critical fluctuations associated with the second order phase transition 
can develop in a rapidly expanding system produced in a relativistic heavy-ion 
collison because of the critical slowing down effect \cite{Ber}. In my opinion,
more promissing strategy would be to search for a first order phase transition 
which should have much more spectacular manifestations \cite{Mish0} discussed below. 
It is interesting to note that, under certain
non-equilibrium conditions, a first order transition is also predicted
for symmetric quark-antiquark matter with zero net baryon density \cite{Sat}.

A striking feature of central heavy-ion collisions at high energies, confirmed
in many experiments (see e.g. \cite{reisdorf,Xu}), is a very strong collective
expansion of matter at later stages of the reaction. This process looks like an
explosion with the matter flow velocities comparable with the speed of light.
The applicability of equilibrium concepts for
describing phase transitions under such conditions becomes questionable
and one should expect strong non-equilibrium effects
\cite{Mish1,Sca,Dum}. In this talk I demonstrate that non-equilibrium phase
transitions in rapidly expanding matter can lead to interesting
phenomena which, in a certain sense, are even easier to observe.

\section{Effective thermodynamic potential}

To make the discussion more concrete, in this talk I adopt a picture of the
chiral phase transition predicted by the linear sigma-model with constituent
quarks \cite{Sca1, Moc}. Then the mean chiral field $\Phi=(\sigma,{\bf
\pi})$ serves as an order parameter. The constituent quark mass is generated 
by interaction with the sigma field, $m=g\sigma$, where $g$ is a corresponding 
coupling constant. The effective thermodynamic potential $\Omega(\Phi;T,\mu)$ 
depends, besides $\Phi$, on temperature $T$ and quark chemical potential 
$\mu=\mu_B/3$. This model respects chiral symmetry which
is spontaneously broken in the vacuum, where the sigma field has a nonzero 
expectation value, $\langle\sigma\rangle=f_{\pi}$, $\langle{\bf \pi}\rangle=0$. 
It is important for our discussion below that the model predicts a phase
diagram on the $(T,\mu)$ plane with a critical point at (T=100 Mev. $\mu$=207 MeV)
and a first order phase transition line at lower T and $\mu$.
A schematic behaviour of $\Omega(T,\mu;\Phi)$ as a function of the order
parameter field $\sigma$ at $\pi=0$ is shown in Fig. 1.
The minima of $\Omega$ determine the stable or metastable states of
matter under the condition of thermodynamical equilibrium, where the pressure
is $P=-\Omega_{min}/V$.
The curves from bottom to top correspond to homogeneous matter at 
different quark chemical potentials and fixed temperature $T=0$.
The dash-dotted curve corresponds to the first order phase transition
point (two equal minima separated by a potential barrier). Two dashed 
curves show the thermodynamic potential at upper and lower spinodal points,
where one of the minima disappears. The range of thermodynamic parameters where 
two phases, one stable and one metastable, may exist simulteneously is 
constrained by these two curves. The critical point would correspond to the 
situation when two minima fuse and the barrier disappears. This situation is 
illustrated in Fig.~2 showing the thermodynamic potential at different 
temperatures and fixed chemical potential $\mu_c=207$ MeV. 
This model reveals a rather weak first order phase transition, although some 
other models \cite{Jac,Raj} predict a stronger transition. The discussion below
is quite general.

\begin{figure*}[htp!]
\centerline{\psfig{file=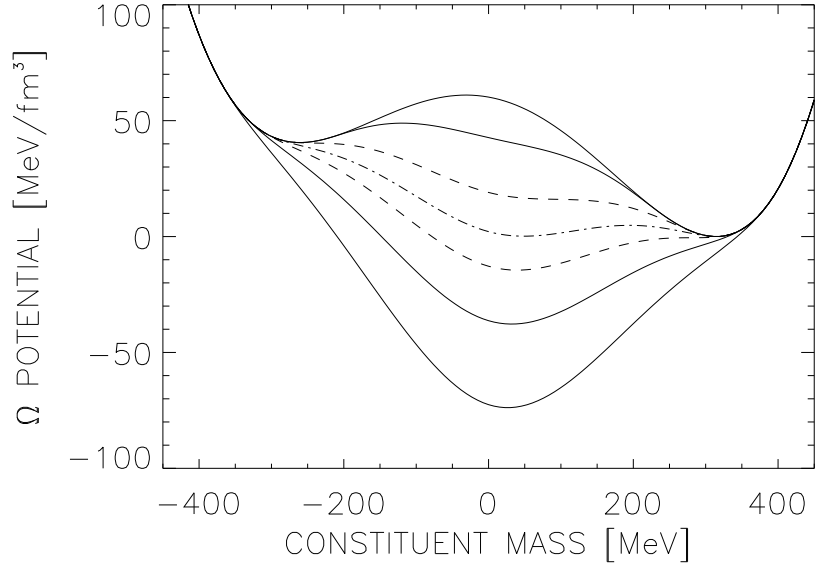,height=7.5cm}}
\caption{The thermodynamical potential $\Omega$ for the sigma model at $T=0$ and 
quark chemical potentials (starting from the top):$\mu=[0,225,279,306,322,345,375]$ MeV.}
\label{potmu}
\end{figure*}  

\begin{figure*}[htp!]
\centerline{\psfig{file=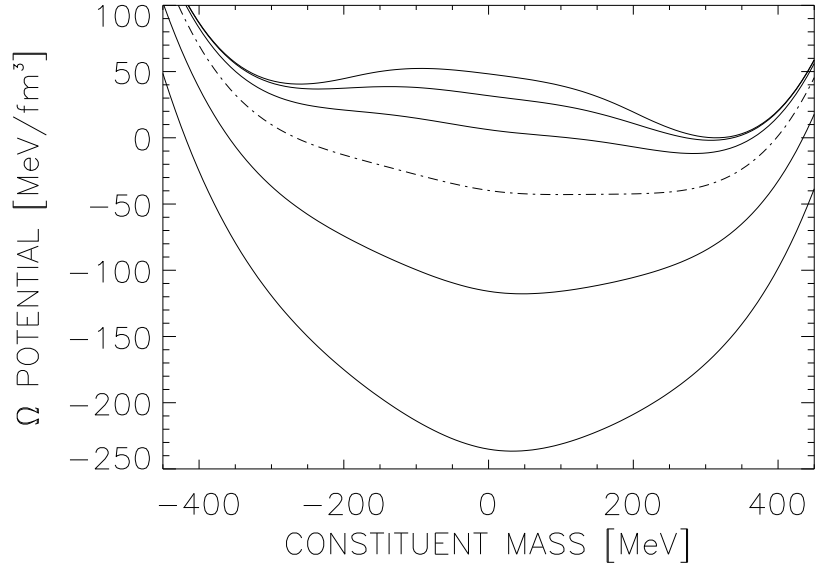,height=7.5cm}}
\caption{The thermodynamical potential $\Omega$ for the sigma model
for the sigma model at $\mu$ fixed to 207 MeV and temperatures 
(starting from the top):$T=[0,50,75,100,125,150]$ MeV.} 
\label{potcrit}
\end{figure*}

One can plot a family of curves for the $(T,\mu)$ values
corresponding to an isentropic expansion of matter. Qualitatively the potential 
curves look similar to the ones depicted in Fig.~1.
Assume that at some early stage of the reaction the thermal equilibrium is
established, and partonic matter
is in a ``high energy density'' phase Q (lowest curve). This state corresponds to the
absolute minimum of $\Omega$ with the order parameter
close to zero, $\sigma\approx 0$, ${\bf \pi}=0$, and chiral symmetry
restored. Due to a very high internal pressure, Q matter will
expand and cool down.
At some stage a metastable minimum appears in $\Omega$
at a finite value of $\sigma$ corresponding to a ``low energy density''
phase H (lower dashed curve), in which chiral symmetry is spontaneously 
broken. At some later time, the critical line in the ($T,\mu$) plane is crossed
where the Q and H minima have equal depths, i.e. $P_{\rm H}=P_{\rm Q}$
(dot-dashed curve). At later times the H phase
becomes more favorable, but the two phases are still
separated by a potential barrier. At certain stage the minimum corresponding 
to the Q phase dissapears (upper dashed curve). The dashed curves separate the 
regions in the phase diagram where one of the phases is unstable (spinodal points).
If the expansion of the Q phase continues until the barrier vanishes, 
the system will find itself in an absolutely unstable state at
a maximum of the thermodynamic potential. Therefore, it will
freely roll down into the lower energy state corresponding to the H phase.
This situation is known as spinodal instability. As shown e.g. in ref. 
\cite{Cse}, the characteristic time of the "rolling down" process is 
relatively short, of about 1 fm/c. 

As well known, a first order phase transition proceeds through the
nucleation process. According to the standard theory of homogeneous nucleation
\cite{Kap}, supercritical bubbles of the H phase can appear only below the critical
line, when $P_H>P_Q$. Then the critical radius for growing bibbles is 
$R_c=2\gamma/(P_H-P_Q)$, where $\gamma$ is the interface energy per unit area
(surface tention).
In rapidly expanding matter the nucleation picture might be very different.
As argued in ref. \cite{Mish0}, the phase separation in this case can begin as
early as the metastable H state appears in the thermodynamic potential, and a
stable interface between the two phases may exist. An appreciable amount of
nucleation bubbles and even empty cavities may be created already above the
critical line. They are stabilized by the collective expansion of matter.

The bubble formation and growth will also continue below the critical line.
Previously formed bubbles will now grow faster due to increasing pressure
difference, $P_{\rm H}-P_{\rm Q}>0$, between the two phases.
It is most likely  that the conversion of Q matter on the bubble boundary
is not fast enough to saturate the H phase. Therefore, a fast expansion
may lead to a deeper cooling of the H phase inside the bubbles compared to
the surrounding Q matter. Strictly speaking, such a system cannot be
characterized by the unique temperature. At some stage the H bubbles
will percolate, and the topology of the system will change to isolated
regions of the Q phase (Q droplets) surrounded by the undersaturated
vapor of the H phase.

\section{Fragmentation of a metastable phase}

The characteristic droplet size can be estimated by applying the energy
balance consideration, proposed by Grady \cite{Gra,Hol} in the study of
dynamical fragmentation of fluids. The idea is that the fragmentation
of expanding matter is a local process minimizing the sum of surface and
kinetic (dilational) energies per fragment volume. As shown in ref. \cite{nn97},
this prescription works fairly well for the multifragmentation of expanding
nuclear systems produced in intermediate-energy heavy-ion reactions, where the
standard~statistical~approach~fails.

Let us consider first an isotropically expanding system where 
the collective velocity field follows locally the Hubble law
\begin{equation}
     v(r)=H\cdot r~,
\end{equation}
where $H$ is a Hubble constant. 
Since there is no preferred direction, the droplets of the Q phase will have 
a more or less spherical shape.
The characteristic radius of the droplet, $R$, can be estimated as follows
\cite{Gra}. The energy $\Delta E$ associated with a Q droplet embedded in
a background of the H phase is represented as the sum of three terms,
\begin{equation} \label{balance}
\Delta E= E_{bulk}+E_{kin}+E_{sur}.
\end{equation}
The bulk term is simply equal to $\Delta{\cal E}V$ where $\Delta{\cal E}=
{\cal E}_Q-{\cal E}_H$ is the energy density difference
between the two bulk phases and $V\propto R^3$ is the volume of the droplet.
The second term is the collective kinetic energy of the droplet expansion
with respect to its center of mass, 
\begin{equation}
E_{kin}=\frac{1}{2}\int {\cal E}v^2(r) dV=
\frac{3}{10}\Delta{\cal E}V H^2 R^2.
\end{equation} 
The last term in eq. (\ref{balance}) is the 
interface energy, $E_{sur}=\gamma S$, which is parametrized in terms of the 
effective surface tension $\gamma$ and the surface area $S\propto R^2$. 
Grady's argument is that the redistribution of matter is a local process 
minimizing the energy per droplet volume, $\Delta E/V$. Then the bulk contribution
does not depend on $R$, and the minimization condition constitutes the balance
between the collective kinetic energy and interface energy. This leads to the 
optimum radius of a droplet
\begin{equation}  \label{R}
R^*=\left(\frac{5\gamma}{\Delta{\cal E}H^2}\right)^{1/3}.
\end{equation}
It is worth noting that the collective kinetic energy acts here as an 
effective long-range potential similar to the Coulomb potential in nuclei.

At ultrarelativistic collision energies associated with RHIC and LHC 
experiments, the expansion of partonic matter will be very anisotropic with 
its strongest component along the beam direction \cite{Bjo}. Clear 
indications of such an anisotropy
are seen already at SPS energies (see \cite{Xu}). It is natural 
to think that in this case the inhomogeneities associated with the 
phase transition will rearrange into pancake-like slabs of Q matter 
embedded in a dilute H phase. The characteristic width of the slab, $2L$, 
can be estimated in a similar way and the resulting expression for $L^*$ 
differs from Eq. (\ref{R}) only by a geometrical factor (3 instead of 5 in 
parentheses). Generally, the faster is the expansion, the smaller are the 
fractures. Of course, at a later time the Q droplets will further fragment 
in the transverse direction due to the standard nucleation process.

As Eq. (\ref{R}) indicates, the droplet size depends 
strongly on $H$. When expansion is slow (small $H$) the droplets are big.
Ultimately, the process may look like a fission of a cloud of plasma.
But fast expansion should lead to very small droplets. This state of matter 
is very far from thermodynamical equilibrium, particularly because the H phase 
is very dilute. One can say that the metastable Q matter is torn apart by a 
mechanical strain associated with the collective expansion. 
This has a direct analogy with the fragmentation of pressurized fluids 
leaving nozzles. In a similar way, splashed water makes droplets which 
have nothing to do with the liquid-gas phase transition.

The driving force for expansion is the pressure gradient, 
$\nabla P=
\equiv c_s^2\nabla {\cal E}$,
which depends crucially on the sound velocity in the matter, $c_s$.
Here we are interested in the expansion rate of the partonic phase which 
is not directly observable.
In the vicinity of the phase transition, one may expect a ``soft point'' 
\cite{Shu,Ris} where the sound velocity is smallest and the ability of 
matter to generate the collective expansion is minimal. If the initial state 
of the Q phase is close to this point, its subsequent expansion will be slow.
Accordingly, the droplets produced in this case will be big. When moving away 
from the soft point, one would see smaller and smaller droplets. For numerical 
estimates we choose two values of the Hubble constant: $H^{-1}$=20 fm/c to 
represent the slow expansion from the soft point and $H^{-1}$=6 fm/c for the 
fast expansion.  

One should also specify two other parameters, $\gamma$ and 
$\Delta{\cal E}$. The surface tension $\gamma$ is a subject of debate at 
present. Lattice simulations indicate that at the critical point it could 
be as low as a few MeV/fm$^2$. However, for our non-equilibrium scenario, 
more appropriate values are closer to 10-20 MeV/fm$^2$ which follow from 
effective chiral models. As a compromise, the value $\gamma=10$ MeV/fm$^2$ is 
used below. Bearing in mind that nucleons and heavy mesons are the smallest 
droplets of the Q phase, one can take $\Delta{\cal E}=0.5$ GeV/fm$^3$, i.e. 
the energy density inside the nucleon. Then one gets $R^*$=3.4 fm for 
$H^{-1}$=20 fm/c and $R^*$=1.5 fm for $H^{-1}$=6 fm/c. 
As follows from eq. (\ref{R}), for a spherical droplet $V\propto 1/\Delta{\cal E}$,
and in the first approximation its mass, 
\begin{equation}
M^*\approx \Delta{\cal E}V=\frac{20\pi}{3}\frac{\gamma}{H^2},
\end{equation} 
is independent of $\Delta{\cal E}$. For two values of $R^*$ given above the 
mass is $\sim$100 GeV and $\sim$10 GeV, respectively. The pancake-like 
droplets could be heavier due to their larger transverse size. 
Using the minimum information principle one can show \cite{Hol,nn97} that the 
distribution of droplets should follow an exponential law, $\exp{\left(-{M 
\over M^*}\right)}$. Thus, about 2/3 of droplets have masses smaller than
$M^*$, but with 1$\%$ probability one can find droplets as heavy as $5M^*$. 

\section{Observable manifestations of quark droplets}

After separation, the droplets recede from each other according to the 
Hubble law, like galaxies in expanding Universe. Therefore, their c.m. 
rapidities are in one-to-one correspondence with their spatial positions.
One may expect that they are distributed more or less uniformly between the target 
and the projectile rapidities. On this late stage it is unlikely that 
the thermodynamical equilibrium is re-established 
between the Q and H phases or within the H phase alone. If this were to 
happen, the final H phase would be uniform, and thus
there would be no traces of the droplet phase in the final state. 

The final fate of individual droplets depends on their sizes and on 
details of the equation of state. Due to the additional Laplace pressure, 
$2\gamma/R$, the residual expansion of individual droplets will slow down. The 
smaller droplets may even reverse their expansion and cooling to shrinking 
and reheating. Then, the conversion of Q matter into H phase may proceed 
through the formation of the imploding deflagration front \cite{Ris,Dig}.
Bigger droplets may expand further until they enter the region of
spinodal instability At this stage the difference 
between 1-st and 2-nd order phase transitions or a crossover is 
insignificant. Since the characteristic ``rolling down'' time is rather short, 
$\sim 1$ fm/c, the Q droplets will be rapidly converted into the non-equilibrium 
H phase. In refs. \cite{Mish1,Dum} the evolution of individual droplets was 
studied numerically within a 
hydrodynamical approach including dynamical chiral fields (Chiral Fluid Dynamics).  
It has been demonstrated that the energy released at the spinodal decomposition 
can be transferred directly into the collective oscillations of the ($\sigma,{\bf \pi}$) 
fields which give rise to the soft pion radiation. One can also expect 
the formation of Disoriented Chiral Condensates (DCC) in the voids between 
the Q droplets. 

An interesting possibility arises if the metastable
Q phase has a point of zero pressure. In particular, this is the case for the 
MIT bag model equation of state at temperatures only slightly below $T_c$
\cite{Kap1}. In this case the droplets might be in mechanical equilibrium 
with the surrounding vacuum ($P_H\approx$0), like atomic nuclei or water 
droplets. The equilibrium condition is 
\begin{equation}
P_Q=\frac{\nu_Q}{2\pi^2}\left[\frac{7\pi^4}{180}T^4+\frac{\pi^2}{6}T^2\mu^2
+\frac{1}{12}\mu^4\right]-B=\frac{2\gamma}{R},
\end{equation}
where $\nu_Q=12$ is the degeneracy factor for massless u and d quarks 
(the gluon contribution is omitted here), and $B$ is a bag constant.
The evolution is then governed by the evaporation of hadrons from the surface
(see also the discussion in Ref. \cite{Alf}). 
One can speculate about all kinds of exotic objects, like e.g. strangelets, 
glueballs, formed in this way. The possibility of forming "vacuum bubbles',
i.e. regions with depleted quark and gluon condensates, was discussed in ref. 
\cite{Mish1}. All these interesting possibilities deserve further study and 
numerical simulations.


\begin{figure*}[htp!]
\centerline{\psfig{file=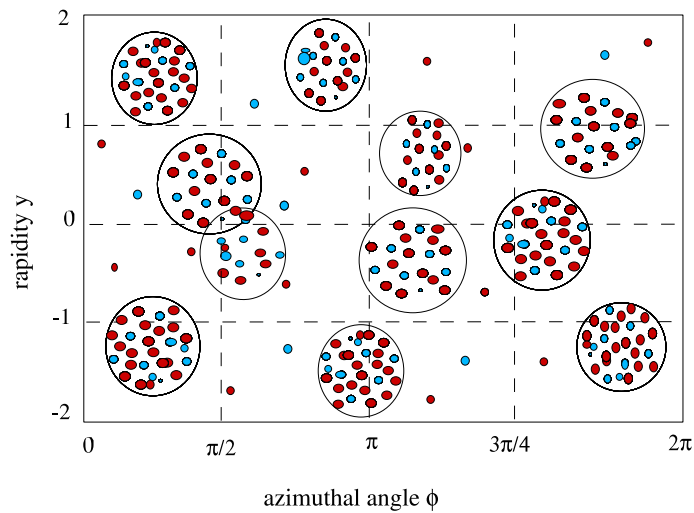,height=7.5cm}}
\caption{Schematic view of the momentum space distribution of secondary 
hadrons produced from an ensemble of droplets. Each droplet emtts hadrons 
(mostly pions) within a rapidity interval $\delta y\sim1$ and azimuthal 
angle spreading of $\delta\phi\sim 1$.} 
\label{potcrit}
\end{figure*}

After separation the QGP droplets recede from each other according to the
global expansion, pre\-do\-mi\-nant\-ly along the beam direction.
Hence their center-of-mass rapidities $y_i$ are in one-to-one correspondence
with their spatial positions.
Presumably $y_i$ will be distributed more or less evenly between
the target and projectile rapidities.
Since rescatterings in the dilute H phase are rare, most hadrons produced
from individual droplets will go directly into detectors. This may explain why 
freeze-out parameters extracted from the hadronic yields are always very close 
to the phase transition boundary \cite {Bra}.    

In the droplet phase the mean number of produced hadrons in a given rapidity 
interval is
\begin{equation} 
\langle N\rangle=\sum\limits_i^{N_D}\overline{n_i}=\langle n\rangle\langle N_D\rangle~,
\end{equation}
where $\overline{n_i}$ is the mean multiplicity of hadrons emitted from a 
droplet i, $\langle n\rangle$ is the average multiplicity per droplet and 
$\langle N_D\rangle $ is the mean number of droplets produced in this interval.
If droplets do not overlap in the rapidity space, each
droplet will give a bump in the hadron rapidity distribution around its
center-of-mass rapidity $y_i$ \cite{Cse}. In case of a Boltzmann spectrum the
width of the bump will be $\delta \eta \sim \sqrt{T/m}$, where $T$ is the
droplet temperature and $m$ is the particle mass. At $T\sim 100$ MeV this
gives $\delta \eta \approx 0.8$ for pions and $\delta \eta \approx 0.3$ for nucleons.
These spectra might be slightly modified by the residual expansion of droplets.
Due to the radial expansion of the fireball the droplets should also be well 
separated in the azimuthal angle. The characteristic angular spreading of pions
produced by an individual droplet is determined by the ratio of the thermal 
momentum of emitted pions to their mean transverse momentum, $\delta\phi\approx 
3T/\langle p_{\perp}\rangle\sim$ 1. 
The resulting phase-space distribution of hadrons in a single event will be a 
superposition of contributions from different Q droplets superimposed on a more
or less uniform background from the H phase. Such a distribution is shown schemmatically 
in Fig.~3. It is obvious that such inhomogeneities (clusterization) in the 
momentum space will reveal strong non-statistical fluctuations.
The fluctuations will be more pronounced if primordial droplets are big,
as expected in the vicinity of the soft point. If droplets as heavy as 100 GeV are 
formed, each of them will emit up to $\sim$200 pions within a narrow rapidity and 
angular intervals, $\delta \eta \sim 1$, $\delta\phi\sim 1$. If only a few droplets 
are produced in average per unit rapidity, $N_D\gtrsim 1$, 
they will be easily resolved and analyzed. On the other hand, the fluctuations 
will be suppressed by factor $\sqrt{N_D}$ if many small droplets shine in the same 
rapidity interval.

\section{Anomalous multiplicity fluctuations}

For our discussion below we consider a more general case of the droplet mass 
distribution when masses follow a gamma-distribution
\begin{equation} \label{md}
w_k(M)=\frac{b}{\Gamma(k)}\left(bM\right)^{k-1}\exp{(-bM)}~,
\end{equation}
which is normalized for $0\leq M\leq\infty$.
The mean mass and its standard deviation are expressed through the parameters 
$k$ and $b$ as
\begin{equation}
<M>=\frac{k}{b}~,~\sigma_M=\frac{\sqrt{k}}{b}=\frac{<M>}{\sqrt{k}}~.
\end{equation} 
These expressions show that quantity $1/\sqrt{k}$ gives the relative scale of 
fluctuations of $M$ around $<M>$. It should be stressed that the gamma-distribution 
(\ref{md}) drops at large $M$ less rapidly than a corresponding gaussian distribution.

One can easily calculate the combined multiplicity distribution produced by
the ensemble of many droplets. Let us assume that the normalized mass distribution
of droplets is $w_k(M)$ and that each droplet emits hadrons according to the Poisson
law, $p_{\overline{n}}(n)$, with the mean multiplicity proportional to the droplet mass,
$\overline{n}=M/\langle E_{\pi}\rangle$ (for pions $\langle E_{\pi}\rangle\approx 3T\sim$
 0.5 GeV). Then the combined distribution is given by the convolution of the two,
\begin{equation}
P_k(N)=\int_0^M dM w_k(M) p_{\overline{n}}(N).
\end{equation}
For the gamma-distribution (\ref{md}) one can perform explicit analytical calculations.
It is remarkable that the resulting distribution is a famous Negative Binomial 
Distribution (NBD)
\begin{equation} \label{NBD}
P_k(N)=\frac{(N+k-1)!}{N!(k-1)!}\frac{\left(\frac{\langle N\rangle}{k}\right)^N}
{\left(1+\frac{\langle N\rangle}{k}\right)^{N+k}}~.
\end{equation}
In a limiting case of the exponential mass distribution (k=1) the 
combined distribution is simply given by
\begin{equation}
P_1(N)=\frac{1}{\langle N\rangle}\left(\frac{\langle N\rangle}{1+
\langle N\rangle}\right)^{N+1},
\end{equation}
where $\langle N\rangle =\langle \overline{n}\rangle=\alpha\langle M\rangle$
is the mean total multiplicity.  

It is convenient to characterize the fluctuations by the scaled variance
\begin{equation}
\omega_N\equiv \frac{\langle N^2\rangle-\langle N\rangle^2}{\langle N\rangle}~.
\end{equation}
Its important preperty is that $\omega_N=1$ for the Poisson distribution, and therefore 
any deviation from unity will signal a non-statistical emission mechanism.
For the NBD, eq. (\ref{NMD}), one easily finds $\omega_N=1+\langle N\rangle/k$. 
As shown in ref. \cite{Bay}, for an ensemble of emitting sources (droplets) 
$\omega_N$ can be expressed in a simple form, 
$\omega_N=\omega_n+\langle n\rangle\omega_D$, 
where $\omega_n$ is an average multiplicity fluctuation in a single droplet,
$\omega_D$ is the fluctuation in the droplet size distribution and $\langle n\rangle$ 
is the mean multiplicity from a single droplet. Since $\omega_n$ and $\omega_D$ 
are typically of order of unity, the fluctuations from the multi-droplet emission
are enhanced by the factor $\langle n\rangle$. According to the picture of
a first order phase transition advocated above, this enhancement factor could
be as large as 10$^2$. It is clear that the nontrivial structure of the hadronic 
spectra will be washed out to a great extent when averaging over many events. 
Therefore, more sophisticated methods of the event sample analysis should be 
applied. As demonstrated below, the simplest one is to  search for non-statistical
fluctuations in the hadron multiplicity distributions measured in a varied
rapidity bin. 

\begin{figure*}[htp!]
\centerline{\psfig{file=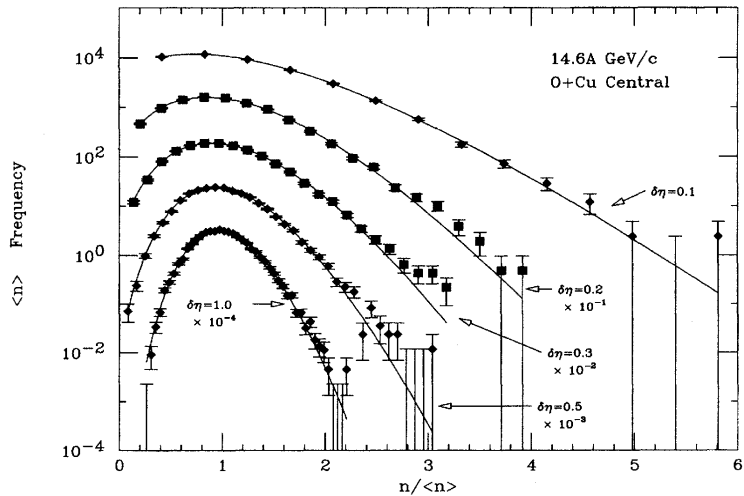,height=7.5cm}}
\caption{Event-by-event multiplicity distributions in $\delta\eta$ intervals 
0.1, 0.2, 0.3, 0.5, 1.0 measured by E-802 Collaboration for central $^{16}$O+Cu 
collisions at lab energy 14.6 AGeV \cite{Abb}. The data for each interval 
are plotted as a function of $n/\langle n\rangle$ and scaled by $\langle n\rangle$, 
the mean multiplicity in the interval. Each successive distribytion has been normalized 
by the factor indicated in the figure. The shape evolvs from almost Gaussian  
($\delta\eta=1.0$) to nearly exponential  ($\delta\eta$=0.1). }
\label{mult1}
\end{figure*}  


The event-by-event multiplicity fluctuations observed so far in heavy-ion experiments 
do not reveal any anomalous enhancement. Fig.~4 shows an example of the multiplicity 
distributions in varying pseudorapidity intervals measured for O+Cu central collisions
at AGS \cite{Abb}. These distributions can very well be fitted  
by the NBD where parameters k$(\delta\eta)$ and $\langle N(\delta\eta)\rangle$ 
follow a linear relationship with nonzero intercept $k(0)$. The data show 
an increase in scaled variances of about 15$\div$30\% over 1. Apparantly such
moderate deviations can be explained by ordinary reasons, not related to a 
phase transition. 
An interseting observation has been made by the NA49 collaboration \cite{Ryb},
which has found non-monotonic behaviour of $\omega_N$ as a function of the 
projectile participant number $N_p$. But in this case too the actual values of
$\omega_N$ are only of about 2, and most likely they can be explained by the 
fluctuations in the number of target participants \cite{Kon}.   

It should be noted that the NBD fits were first used to 
describe the multiplicity distributions in high-energy $pp$ and $p\bar{p}$ 
collions (see e.g. ref.\cite{Cur}). They are consistent with the so called 
KNO scaling \cite{KNO}.

\section{Conclusions}

\begin{itemize}
\item A first order phase transition in rapidly expanding matter should proceed
through the nonequilibrium stage when a metastable phase splits into droplets
whoose size is inversly proposrtional to the expansion rate.
The primordial droplets should be biggest in the vicinity of a soft point when
the expansion is slowest.


\item Hadron emission from droplets of the quark-gluon plasma should lead to
large nonstatistical fluctuations in their rapidity and azimuthal spectra, 
as well as in multiplicity distributions in a given rapidity window. 
The hadron abundances may reflect directly the chemical composition in 
the plasma phase.

\item To identify the phase transition threshold the measurements should be
done at different collision energies.
The predicted dependence on the expansion rate and the reaction geometry
can be checked in collisions with different ion masses and
impact parameters.

\item If the first order deconfinement/chiral phase transition is only possible 
at finite baryon densities, one should try to identify it by searching for the 
anomalous fluctuations in the regions of phase space characterized by a large 
baryon chemical potential. These could be the nuclear fragmentation regions 
in collisons with very high energies (high-energy SPS, RHIC, LHC) or the central 
rapidity region (AGS, low-energy SPS, future GSI facility FAIR).

\end{itemize}

The author is grateful to L.M. Satarov, M.I. Gorenstein and M. Gazdzicki for many 
fruitful discussions. This work was supported partly by the RFBR grant 05-02-04013 
(Russia).

\vspace{-0.3cm}

\end{document}